\documentclass[aip,apl,twocolumn,amsmath,amssymb,floatfix,reprint,square,comma,numbers]{revtex4}
\usepackage[dvipdfm,colorlinks=true,linkcolor=blue]{hyperref}
\usepackage{graphicx}
\usepackage{makeidx}
\usepackage[normalem]{ulem}
\usepackage{braket}
\DeclareMathOperator{\Tr}{Tr}
\makeindex

\begin{document}
\begin{@twocolumnfalse}
\title{Reducing intrinsic decoherence in a superconducting circuit by quantum error detection}

\author{Y. P. Zhong}
\affiliation{Department of Physics, Zhejiang University, Hangzhou 310027 China}
\author{Z. L. Wang}
\affiliation{Department of Physics, Zhejiang University, Hangzhou 310027 China}
\author{John M. Martinis}
\affiliation{Department of Physics, University of California, Santa Barbara, CA 93106 USA}
\author{A. N. Cleland}
\affiliation{Department of Physics, University of California, Santa Barbara, CA 93106 USA}
\author{A. N. Korotkov}
\affiliation{Department of Electrical Engineering, University of
California, Riverside, CA 92521 USA}
\author{H. Wang}
\email{hhwang@zju.edu.cn}
\affiliation{Department of Physics, Zhejiang University, Hangzhou 310027 China}

\date{\today}

\begin{abstract}
A fundamental challenge for quantum information processing is reducing the impact of environmentally-induced errors. Quantum error detection (QED) provides one approach to handling such errors, in which errors are rejected when they are detected.
Here we demonstrate a QED protocol based on the idea of quantum un-collapsing, using this protocol to suppress energy relaxation due to the environment in a
three-qubit superconducting circuit. We encode quantum information in a target qubit, and use the other two qubits to detect and reject errors caused by energy relaxation. This protocol improves the storage time of a quantum state by a factor of roughly three, at the cost of a reduced probability of success.
This constitutes the first experimental demonstration of an algorithm-based improvement in the lifetime of a quantum state stored in a qubit.
\end{abstract}

\maketitle
\end{@twocolumnfalse}

Superconducting quantum circuits are very promising candidates for
building a quantum processor, due to the combination of good qubit
performance and the scalability of planar integrated
circuits~\cite{Clarke2008, You2005, Devoret2013, Sun2010, Niskanen2007, Mariantoni2011,
Paik2011, Rigetti2012, Barends2013, Abdumalikov2013}. In addition to recent, very
significant improvements in the materials and qubit geometries in such circuits,
external control and measurement protocols are being developed to
improve performance. This includes the use of dynamical
decoupling~\cite{Bylander2011}, and preliminary experiments~\cite{Reed2012} with
quantum error correction codes, which allow the removal of
artificially-induced errors~\cite{Shor1995, Leung1997, Yao2012,
Schindler2011, Reed2012}. To date, however, there has been
little experimental progress in control sequences that reduce a
significant source of qubit error, energy dissipation due to the
environment.

Quantum error detection (QED)~\cite{Knill2005, Korotkov2012} provides an
alternative, albeit non-deterministic approach to handling errors, avoiding some of the
complexity of full quantum error correction by simply rejecting
errors when they are detected. QED has been predicted to
significantly reduce the impact of energy relaxation in
qubits~\cite{Korotkov2012}, one of the dominant sources of error in
superconducting quantum circuits~\cite{You2005, Clarke2008,
Devoret2013}.
Here we demonstrate a QED
protocol in a circuit comprising a target qubit entangled with two
ancilla qubits, using a variant of the quantum un-collapsing
protocol that combines a weak measurement with its
reversal~\cite{Korotkov2006, Katz2008, Zubairy2009, Korotkov2010}. We use this protocol to successfully extend the intrinsic lifetime of a quantum state by a factor of about three. A
somewhat similar protocol has been demonstrated with photonic
qubits, but only to suppress intentionally-generated
errors~\cite{Kim2011b}.

\begin{figure*}[t]
  \includegraphics[width=5.0in, clip=True]{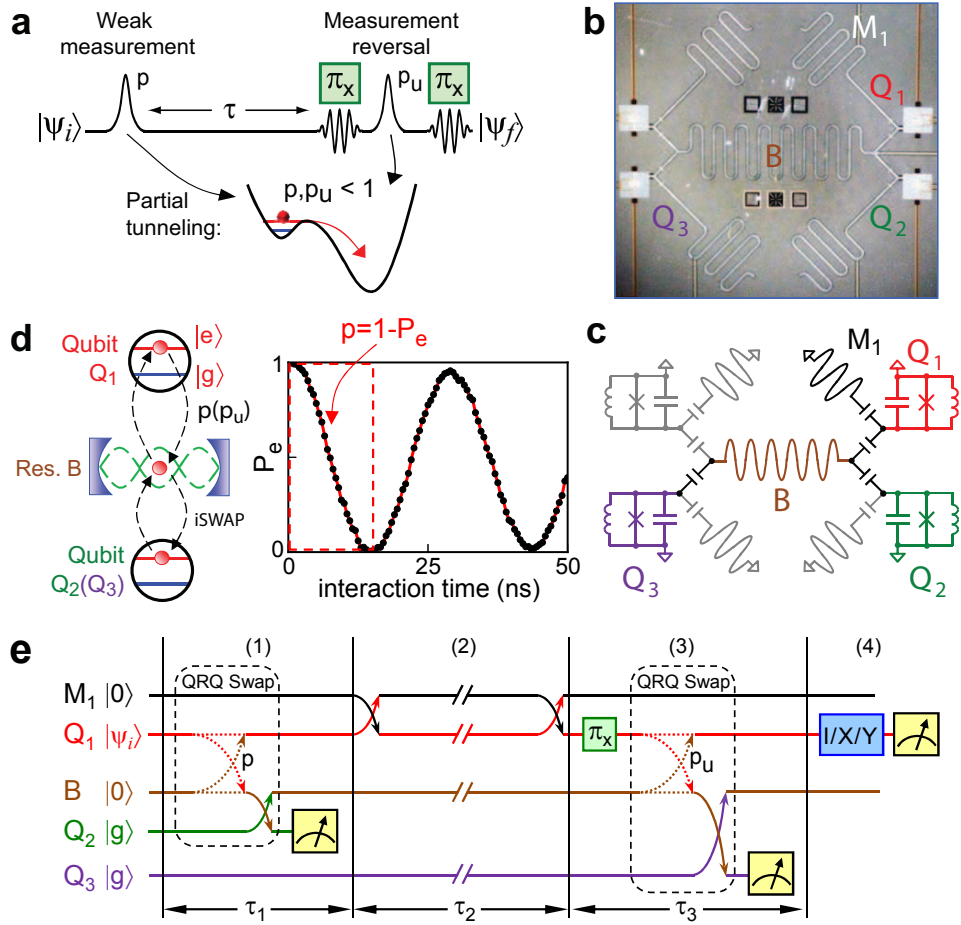}
  \caption{\label{fig1}\footnotesize{\textbf{Device geometry and
un-collapsing protocol used for QED.}
  \textbf{a},~Quantum un-collapsing protocol in the phase
qubit~\cite{Katz2008,Korotkov2010}. Top: Pulse sequence, where the
weak measurement with strength $p$ is followed by a delay (storage
time) $\tau$, and then the measurement reversal, involving a $\pi_x$
rotation, a weak measurement with strength $p_u$, and a second
$\pi_x$ rotation. Bottom: The delta-like electrical pulses lower the
tunnel barrier for the qubit states on the left of the potential
landscape to allow partial tunneling of the $|e\rangle$ state into
the well on the right.
  \textbf{b-c},~Optical micrograph and simplified schematic of the device. Circuit elements are as labeled; those not used in this experiment are in gray.
  \textbf{d},~Illustration of the qubit-resonator-qubit (QRQ) swap, analogous
to the partial tunneling measurement. Left: Schematic for the
sequential qubit $Q_1$-resonator $B$ swap with swap probability
(measurement strength) $p$ ($p_u$), followed by a full iSWAP between resonator $B$ and qubit
$Q_2$ ($Q_3$). Right: The on-resonance, unit-amplitude
qubit-resonator vacuum Rabi oscillations in the qubit $|e\rangle$
state probability $P_e$ (vertical axis), starting with the qubit in
$|e\rangle$ and resonator in $|0\rangle$. The
measurement strength $p = 1 - P_e$ is set by the interaction time
(horizontal axis).
  \textbf{e},~QED protocol, where we start with $Q_1$ in
$|\psi_i\rangle$, consisting of the following
steps: 1. The first weak measurement is performed using the first
QRQ swap involving $Q_1$-$B$-$Q_2$, with strength $p$. $Q_2$ is
measured immediately, and only null outcomes ($Q_2$ in
$|g\rangle$) are accepted.
2. The state is swapped from $Q_1$ into memory resonator $M_1$ and
stored for a relatively long time $\tau_2$, following which the state is swapped back into
$Q_1$. 3. The weak measurement reversal is performed using a $\pi_x$
rotation on $Q_1$ and a second QRQ swap with strength $p_u$ to qubit
$Q_3$. $Q_3$ is then measured, and only null outcomes ($Q_3$ in
$|g\rangle$) are accepted.
4. The double-null outcomes are analyzed
using tomography of $Q_1$ to evaluate $Q_1$'s final density matrix.
To save time and reduce errors, we do not perform the final $\pi_x$
rotation appearing in the full un-collapsing protocol.}
  }
\end{figure*}

The un-collapsing protocol~\cite{Korotkov2010} we use for QED is
illustrated in Fig.~\ref{fig1}a. Starting with a qubit in a superposition of its ground $|g\rangle$ and excited $|e\rangle$ states,
$|\psi_i\rangle = \alpha |g\rangle + \beta |e\rangle$, a weak
measurement is performed that detects the $|e\rangle$ state with
probability (measurement strength) $p < 1$. In the null-measurement
outcome ($|e\rangle$ state not detected), this produces the
partially collapsed
state $|\psi_1\rangle = \alpha |g\rangle + \beta \sqrt{1-p}
|e\rangle$ (the squared norm equals the outcome probability). The
system is then stored for a time $\tau$, during which it can decay
(``jump'') to the state $|g\rangle$, or remain in the ``no-jump''
state
   $|\psi^{nj}\rangle = \alpha|g\rangle + \beta \sqrt{1-p}\,
   e^{-\Gamma \tau/2} |e\rangle$,
where $\Gamma=1/T_1$ is the energy relaxation rate.
The un-collapsing measurement is then performed, comprising a
$\pi_x$ rotation and a second weak measurement with strength $p_u$,
followed by a final $\pi_x$ rotation that undoes the first rotation.
Only outcomes that yield a second null measurement are kept.
These double-null outcomes give the result $|\psi^j_f \rangle = |g\rangle$ if the system jumped to $|g\rangle$ during the time interval $\tau$, while in the no-jump case, the final state is
\begin{equation}\label{eq2}
    |\psi^{nj}_f \rangle = \alpha \sqrt{1-p_u} \, |g\rangle +
    \beta \sqrt{1-p} \, e^{-\Gamma \tau/2} |e\rangle.
\end{equation}

Remarkably, the final no-jump state is identical to $|\psi_i\rangle$
if we choose $1-p_u = (1-p) e^{-\Gamma\tau}$; the probability of
this (desired) outcome is $P_f^{nj} = \langle \psi_f^{nj}|\psi_f^{nj}\rangle = (1-p)e^{-\Gamma \tau}$, while
the probability of the undesirable jump outcome $|g\rangle$ is
$P_f^{j} = \left|\beta\right|^2 (1-p)^2 e^{-\Gamma\tau}
(1-e^{-\Gamma\tau})$.~\cite{Note2} As the probability $P_f^{j}$
falls to zero more quickly than $P_f^{nj}$  as $p \rightarrow 1$,
increasing the measurement strength $p$ towards 1 results in a high
likelihood of recovering the initial state. This comes at the
expense of a low probability $P_{\rm{DN}}= P_f^{nj}+P_f^j$ of the double-null result.\\

\noindent\textbf{Results}

\noindent The weak measurement in Fig.~\ref{fig1}a is performed by partial
tunneling. We used partial tunneling for the measurement in QED (see
below), but as it consistently yielded low fidelities, we also
developed an alternative, more extensive device and protocol, shown
in Fig.~\ref{fig1}b-d. The device is similar to that in
Ref.~\cite{Lucero2012}, with three phase qubits, $Q_1$, $Q_2$, and $Q_3$,
coupled to a common, half-wavelength coplanar waveguide bus
resonator $B$, with a memory resonator $M_1$ also coupled to
$Q_1$. Relevant parameters are tabulated in the Supplementary
Information.

The alternative partial measurement method is illustrated in Fig.~\ref{fig1}d.
Qubit $Q_1$ is the target, and $Q_2$ and $Q_3$ are ancillae,
entangled with $Q_1$ via the resonator bus $B$, such that a
projective measurement of $Q_2$ or $Q_3$ results in a weak
measurement of $Q_1$. The entanglement begins with a partial swap
between $Q_1$ and the resonator $B$: When qubit $Q_1$, initially in
$|e\rangle$, is tuned to resonator $B$, the probability $P_e$ of
finding the qubit in $|e\rangle$ oscillates with unit amplitude at
the vacuum Rabi frequency~\cite{Hofheinz2009, Wang2013,
Zubairy2011}.  A partial swap with swap probability $p = 1-P_e$ is
achieved by controlling the interaction time, entangling $Q_1$ and
$B$. We then use a complete swap (an ``iSWAP'') between resonator
$B$ and qubit $Q_2$ ($Q_3$), transferring the entanglement, followed
by a projective measurement of $Q_2$ ($Q_3$). In general, we start
with $Q_1$ in $|\psi_i\rangle = \alpha|g\rangle + \beta|e\rangle$
and perform the qubit-resonator-qubit (QRQ) swap, followed by
measurement of the ancilla. A null outcome ($Q_2$ or $Q_3$ in
$|g\rangle$) yields the $Q_1$ state $\alpha|g\rangle + \beta
\sqrt{1-p} |e\rangle$, as with partial tunneling. The swap
probability $p$ is therefore equivalent to the measurement
strength.

Our QED protocol can protect against energy decay of the quantum state. However, as dephasing in
these qubits is an important error source, against which the QED
protocol does not protect, we store the intermediate quantum state
in the memory resonator $M_1$, which does not suffer from dephasing
(as indicated by $T_2 \cong 2 T_1$
for the resonator; see Supplementary Information).

Our full QED protocol is shown in Fig.~\ref{fig1}e, starting with the
initial state of the system as
\begin{equation}\label{eq2B}
    |\Psi_{i}\rangle=(\alpha|ggg\rangle+\beta|egg\rangle)\otimes|00\rangle,
\end{equation}
where $|q_1 q_2 q_3 \rangle$ represents the state of the qubits
$Q_1$, $Q_2$ and $Q_3$, with the ground state $|00\rangle$ of the
$B$ and $M_1$ resonators listed last. In step 1, we use a QRQ swap between $Q_1$, $B$ and $Q_2$ with swap
probability (measurement strength) $p$, followed immediately by
measurement of $Q_2$. This step takes a time $\tau_1$ of up to 15~ns, depending on $p$. A null outcome ($Q_2$ in $|g\rangle$) yields
    $|\Psi_{1}\rangle = \alpha |ggg\rangle |00\rangle
    + \beta \sqrt{1-p} \, |egg\rangle |00\rangle$
(a more precise expression appears in the Supplementary Information). In step 2, we
swap the quantum state from $Q_1$ into $M_1$, wait a relatively long
time $\tau=\tau_2$, during which the state in $M_1$ decays at a rate
$\Gamma=1/T_1$, and we then swap the state back to $Q_1$. In the
no-jump case, the state becomes
    $|\Psi_{2}^{nj}\rangle = \alpha |ggg\rangle |00\rangle
  + \beta \sqrt{1-p} \, e^{-\Gamma\tau_2/2} |egg\rangle |00\rangle$.
We then perform step 3, comprising a $\pi_x$ rotation on $Q_1$ followed by
the second QRQ swap with strength $p_u$, involving $Q_1$, $B$ and
$Q_3$, which takes a time $\tau_3$. $\tau_3$ is between 20 and 35 ns, depending on $p_u$, dominated by the 20 ns-duration $\pi_x$ pulse. $Q_3$ is then measured, with
a null outcome ($Q_3$ in $|g\rangle$) corresponding to
\begin{equation}\label{eq3}
    |\Psi_{f}^{nj}\rangle = (\alpha \sqrt{1-p_u}|e\rangle+\beta
    \sqrt{1-p}\,
    e^{-\Gamma \tau_2/2} |g\rangle)\otimes|gg\rangle|00\rangle.
\end{equation}
We recover the initial state $|\Psi_i\rangle$ if we set $1-p_u = (1-p)e^{-\Gamma \tau_2}$, with the undesired jump cases mostly eliminated by the double-null selection. To shorten the sequence, we do not perform the final
$\pi_x$ rotation, so the amplitudes of $Q_1$'s $|g\rangle$ and $|e\rangle$ states are reversed compared
to the initial state. In step 4, we apply tomography pulses and then measure $Q_1$ to determine its final state, keeping the results that correspond to the double-null outcomes ($Q_2$ and $Q_3$ in $|g\rangle$).

\begin{figure}[t]
  \centering
  \includegraphics[width=3.3in,clip=True]{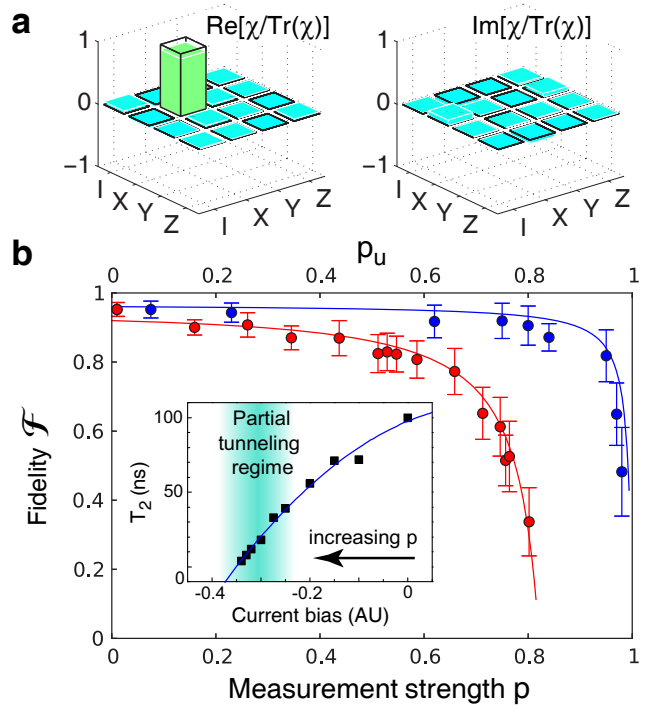}
  \caption{\label{fig2}\footnotesize{\textbf{Fidelity of the uncollapsing protocol
without storage.}
  \textbf{a}, Measured $\chi/\Tr(\chi)$ (bars with color), where $\chi$ is
the non-trace-preserving quantum process tomography matrix for the
sequence in Fig.~\ref{fig1}e excluding step 2, here with
$p=p_u=0.75$. The desired matrix, $\chi^{\textrm{ideal}}$,
corresponds to a $\pi$ rotation about the Bloch sphere $x$ axis
(identified by black frames).
  \textbf{b}, Process fidelity ${\mathcal F}$ for both the three-qubit QRQ-based
un-collapsing (blue circles) and the single-qubit partial-tunneling
version (red circles)~\cite{Katz2008}, both as a function of $p =
p_u$. Error bars represent statistical errors extracted from
repeated measurements. The process fidelity is above 0.9 for $p \leq
0.8$ using the QRQ swaps, while for the partial tunneling scheme it
decreases significantly for $p \geq 0.5$. This decrease is primarily
due to reduction in qubit $T_2$ with measurement current bias, shown
in the inset; partial tunneling occurs in the shaded region. Blue
line is a simulation using $\kappa_1=\kappa_3=0.985$, $\kappa_2=1$,
and $\kappa_\varphi = 0.95$ (see Supplementary Information); the red
line is a guide to the eye.}
  }
\end{figure}

We use quantum process tomography to characterize the performance of
the protocol, starting with the four initial states
$\left\{|g\rangle, |g\rangle-i|e\rangle, |g\rangle+|e\rangle,
|e\rangle\right\}$ and measuring the one-qubit process matrix
$\chi$. As we reject outcomes where $Q_2$ and $Q_3$ are not measured
in $|g\rangle$, the process is not trace-preserving, so the linear
map satisfies $\rho_{f} P_{\rm DN} = \sum_{n,m} \chi_{nm} E_n
\rho_{i} E^{+}_m$, where $\rho_{i}$ and $\rho_{f}$ are the
normalized initial and final density matrices of $Q_1$, and $E_n$ is
the standard Pauli basis $\left\{I, X, Y, Z\right\}$. We define the
process fidelity ${\mathcal F}$ as \cite{Kiesel2005} ${\mathcal
F}=\Tr(\chi^{\textrm{ideal}}\chi)/\Tr(\chi)$, where
$\chi^{\textrm{ideal}}$ corresponds to the desired unitary operation
(here given by $\pi_x$), and the divisor accounts for
post-selection.\cite{Note1}

\begin{figure*}
  \centering
  \includegraphics[width=5.0in,clip=True]{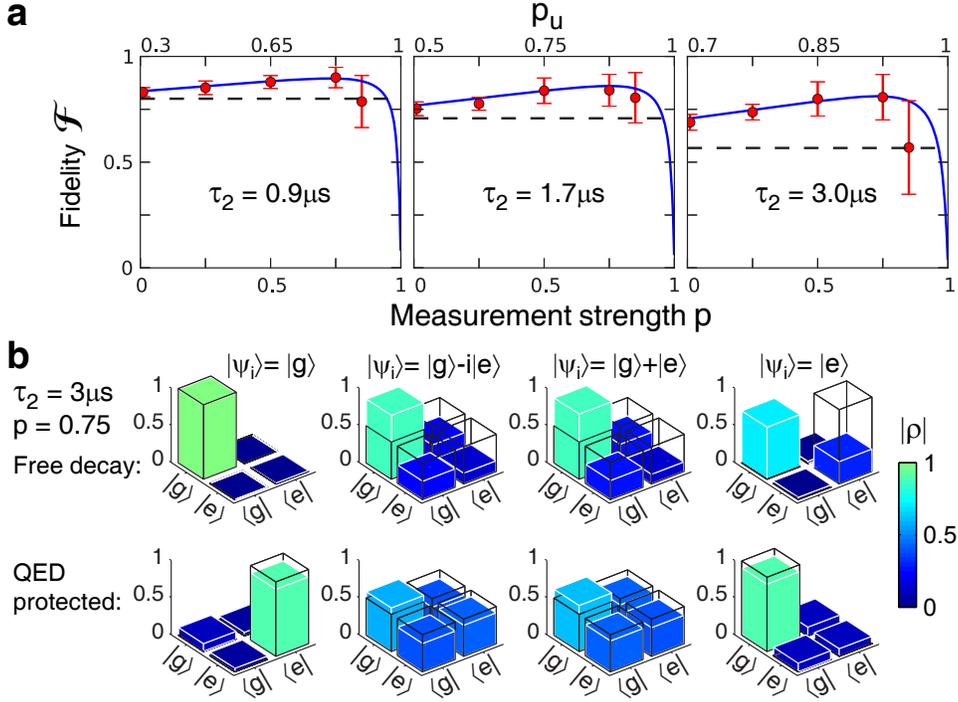}
  \caption{\label{fig3}\footnotesize{\textbf{QED-based quantum state protection
  from energy relaxation.}
  \textbf{a}, Process fidelity ${\mathcal F}$ as a function of measurement strength $p$ for the full QED protocol
for three storage times $\tau_2 = 0.9,~1.7$ and 3 $\mu$s in memory
resonator $M_1$ ($T_1=2.5 \mu$s). The un-collapsing swap probability
$p_u$ is indicated on the top axis (see text). Circles with error
bars are measured data; lines are simulations (see Supplementary
Information). Horizontal dashed lines in each panel give the
free-decay process fidelity; the improvement from QED is most
significant for larger $\tau_2$. The statistical errors increase
with increasing QRQ measurement strength $p$, due to the decrease in
sample size (fewer double-null outcomes); we compensate for dynamic
phases (see Supplementary Information).
  \textbf{b}, Final density matrices (bars with color) without (top row) and with
(bottom row) QED, with $p=0.75$, for the four initial states as
labeled, following a $\tau_2 = 3~\mu$s storage time
($e^{-\tau_2/T_1} = 0.3$). The desired error-free density matrices
are shown by black frames. We only display the absolute values of
the density matrix elements $|\rho|$. Note that the QED-protected
final states differ from the initial state by a $\pi$ rotation.
  }
  }
\end{figure*}

We first tested the process with no storage, entirely omitting step
2 in Fig.~\ref{fig1}e, and choosing $p_u = p$; we also delayed the measurement of $Q_2$ to the end of step 3
to minimize crosstalk (see Methods). Figure \ref{fig2}a
shows the measured $\chi/\Tr(\chi)$ for $p=p_u=0.75$; the calculated
process fidelity is ${\mathcal F}=0.92$.  In Fig.~\ref{fig2}b we
show the measured process fidelity ${\mathcal F}$ as a function of the QRQ
measurement strength $p = p_u$ (blue circles).

We can compare our no-storage un-collapsing fidelity to that obtained
using partial tunneling for the weak measurement of a single qubit
\cite{Katz2008}, shown in Fig.~\ref{fig2}b (red circles). We see
that even though the QRQ-based protocol is more complex, it achieves
much better fidelities for $p \geq 0.5$. This is mostly because of
strong dephasing and two-level state effects~\cite{Wang2013,Sun2010}
during the partial tunneling current pulse (see inset in
Fig.~\ref{fig2}b).

We then tested the full QRQ protocol's ability to protect from energy
decay. The un-collapsing
strength $p_u$ is given by~\cite{Korotkov2010}
$1-p_u=(1-p)\kappa_1\kappa_2/\kappa_3$, where $\kappa_2 =
\exp(-\tau_2/T_1)$ and $\kappa_1$ and $\kappa_3$ are similar energy
relaxation factors for the steps 1 and 3 (here
$\kappa_1\approx\kappa_3\approx0.985$; see Supplementary Information). In Fig.~\ref{fig3}a we
display the measured fidelities for the storage durations $\tau_2 =
0.9,~1.7$ and 3 $\mu$s for the memory resonator with $T_1 = 2.5~\mu$s, compared to simulations using
the pure dephasing factor $\kappa_{\varphi} = 0.95$ (see Ref.~\cite{Korotkov2010} and Supplementary Information). The simulations are in excellent agreement with the data,
and we see a marked improvement in the storage fidelity using QED
over that of free decay (dashed line in each panel).

It is interesting to note that in Fig.~\ref{fig3}a, the process
fidelity is significantly improved even for zero measurement
strength $p=0$ (note that $p_u > 0$), implying that a simpler QED
protocol still provides some protection against energy relaxation.

Another way to test QED is to monitor the evolution of individual
quantum states. In Fig.~\ref{fig3}b we display the final density
matrices measured either without (top row) or with (bottom row) QED,
for four initial states in $Q_1$, with storage in the memory $M_1$
for $\tau_2 \approx 3~\mu$s. Other than for the initial ground state
$|g\rangle$, which does not decay, we see that the QED-protected
states are much closer to the desired outcomes than the free-decay
states (note the $\pi$ rotation). If we look at the off-diagonal
terms in the middle panels, they
have decayed from 0.5 to about 0.4; this decay takes about 1.1
$\mu$s without QED, so the lifetime is increased by $3~\mu{\rm
s}/1.1~\mu{\rm s} \approx 3$. Also, if we look at Fig.~\ref{fig3}a,
the free-decay fidelity at 0.9 $\mu$s (left panel) is about the same
as the maximum QED fidelity at 3.0 $\mu$s (right panel), also giving
a factor of three improvement.

\begin{figure}
  \centering
  \includegraphics[width=3.3in,clip=True]{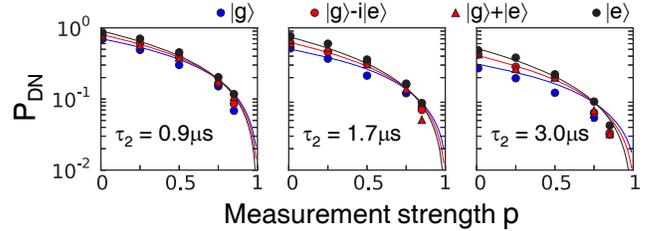}
  \caption{\label{fig4}\footnotesize{\textbf{QED selection probability.} The QED
protocol uses post-selection to reject state decay
 errors. The probability of accepting an
outcome, i.e. the double-null probability $P_{\textrm{DN}}$, falls
with measurement strength $p$. Here we display $P_{\textrm{DN}}$ as a function of $p$, corresponding
to the data in Fig.~\ref{fig3}a, for each value of storage time
$\tau_2$. Lines are predicted by theory.
  }
  }
\end{figure}

The price paid for the lifetime improvement is the small fraction of
outcomes accepted by the QED post-selection, shown in
Fig.~\ref{fig4}. The double-null probability $P_{\rm{DN}}$ decreases
with increasing measurement strength $p$ for
all initial states. A balance must therefore be struck between a
larger $T_1$ improvement, occurring for larger $p$, and a larger
fraction of accepted outcomes, which occurs for smaller $p$.\\

In conclusion, we have implemented a practical QED protocol, based
on quantum un-collapsing, that suppresses the intrinsic energy
relaxation of a quantum state in a superconducting circuit,
increasing the effective lifetime by about a factor of three.
We note that the phase qubits in our design could be
replaced by better-performing qubits~\cite{Barends2013}, on which
real-time quantum non-demolition measurement and feedback control
are feasible\cite{Riste2012,Vijay2012,Devoret2013}. This could enable sufficient coherence for demonstrating a practical fault-tolerant quantum architecture.\\

\noindent\textbf{Methods}

\noindent\footnotesize{\textbf{Readout correction and crosstalk
cancellation.} All data are corrected for the qubit readout
fidelities before further processing. The readout fidelities for
$|g\rangle$ ($F_g$) and $|e\rangle$ ($F_e$) of $Q_1$, $Q_2$, and
$Q_3$ are $F_{1g}=0.95$, $F_{1e}=0.89$, $F_{2g}=0.94$,
$F_{2e}=0.88$, $F_{3g}=0.94$, $F_{3e}=0.91$, respectively. Crosstalk
is another concern when performing QED to protect quantum states. We
read out $Q_2$ immediately after the first QRQ swap in step 1 in
Fig.~\ref{fig1}e to avoid decay in $Q_2$. However, due to
measurement crosstalk in the qubit circuit, this measurement can
result in excitations in resonator $B$; while this does not directly
affect the other qubits, we must reset the resonator prior to the
second QRQ swap. This is done during the storage in the memory resonator, by
performing a swap between $B$ and $Q_3$, and then using a spurious
two-level defect coupled to $Q_3$ to erase the excitation in $Q_3$.
As the storage time in $M_1$ is several microseonds, there is
sufficient time to reset both $B$ and $Q_3$ prior to the second QRQ
swap.

The intermediate reset of $B$ could not be performed when doing the
experiments in Fig.~\ref{fig2}, for which there is no storage interval. To
avoid crosstalk in those measurements, we postponed the measurement
of $Q_2$ until the end of the second QRQ sequence, to step 3 of
Fig.~\ref{fig1}e. The $|e\rangle$ state probability in $Q_2$
drops by about 6\% during this delay time, as estimated from $Q_2$'s
$T_1$. We have corrected for this drop when evaluating the
$Q_2$ measurements for Fig.\ref{fig2}}.

\noindent{\textbf{Acknowledgments}}\\
\noindent{\footnotesize{This work was supported by the National
Basic Research Program of China (2012CB927404), the National Natural
Science Foundation of China (11222437, 11174248, and J1210046), Zhejiang
Provincial Natural Science Foundation of China (LR12A04001), and
IARPA/ARO grant W911NF-10-1-0334 of USA.
H.W. acknowledges supports by Program for New Century Excellent Talents in University (NCET-11-0456) and by
Synergetic Innovation Center of Quantum Information and Quantum Physics.
Devices were made at the UC Santa Barbara Nanofabrication Facility, a part of the NSF funded National Nanotechnology Infrastructure Network.}}\\
\\
\noindent{\textbf{Author contributions}}\\
\noindent{\footnotesize{Y.P.Z., A.N.K., H.W. designed and analyzed the experiment carried out by Y.P.Z. All authors contributed to the experimental set-up and helped to write the paper.}}

\clearpage

\noindent\textbf{\Large\sffamily{Supplementary Information}}

\newcommand{\comments}[1]{}

\renewcommand\thefigure{S\arabic{figure}}
\setcounter{figure}{0}
\renewcommand\theequation{S\arabic{equation}}
\setcounter{equation}{0}
\renewcommand\thepage {S\arabic{page}}
\setcounter{page}{1}
\renewcommand\thetable{S\arabic{table}}
\setcounter{table}{0}

\section{Qubit and resonator parameters used in experiment}

The qubits and resonators used in this experiment were all produced in a multi-layer lithographic process on single-crystal sapphire substrates. The qubits are phase qubits, each consisting of a $2\,\mu\mathrm{m}^2$ Al/AlO$_x$/Al junction in parallel with a $1$ pF Al/a-Si:H/Al shunt capacitor and a 720 pH loop inductance (design values). The resonators are single-layer aluminum coplanar waveguide resonators. We use interdigitated coupling capacitors between the qubits and the resonators. Standard performance parameters of individual elements are listed in Table~\ref{tab1}.

\begin{table}[ht]
  \centering
  \begin{tabular}{c c c c c c c c c c c }
  \hline
  \hline
   &freq.& & $T_1$ && $T_2$ && $T_{SE}$ &&coupling strength\\
   &(GHz)&&(ns)&&(ns)&&(ns)&&(MHz)\\
  \hline
  $Q_1$ & 6.01 & & 580 & & 140 & & 500 & &34.7~($\leftrightarrow B$) \\
  $Q_2$ & 5.90 & & 614 & & 100 & & 510 & &34.1~($\leftrightarrow B$)\\
  $Q_3$ & 5.81 & & 580 & & 150 & & 430 & &33.3~($\leftrightarrow B$)\\
  $B$   & 6.24 & & 3000 & & $\sim$5000 & & $\ast$ & & \\
  $M_1$ & 7.55 & & 2500 & & $\sim$5000 & & $\ast$ & & 56.8~($\leftrightarrow Q_1$) \\
  \hline
  \end{tabular}
  \caption{Operating characteristics for qubits $Q_1$, $Q_2$, $Q_3$,
the bus resonator $B$, and the memory resonator $M_1$. We show the
$|g\rangle-|e\rangle$ splitting frequency for the qubits, the
resonance frequency for the resonators, as well as each element's
measured energy relaxation time $T_1$, Ramsey dephasing time $T_2$,
and spin-echo dephasing time $T_{SE}$. Qubit lifetimes are at the
listed frequencies, and resonator lifetimes are measured using
photon swaps with a qubit; the coupling strengths are from vacuum
Rabi
oscillations.~\cite{Hofheinz2009,Lucero2012,Wang2013}}\label{tab1}
\end{table}

\section{State evolution during the QRQ-based quantum error
detection protocol}

In this section we discuss the state evolution in the actual
experimental protocol, based on the QRQ swaps. We include the
dynamic phases in the analysis but for
simplicity neglect imperfections as well as
decoherence in the unitary operations, while including energy
relaxation during the state storage in the memory resonator (step 2
in Fig.\ 1e of the main text).

Assuming no errors in the preparation of the target qubit $Q_1$, the
initial state of the system prior to step 1 shown in Fig.\ 1e is
[see Eq.\ (2) in the main text]
\begin{equation}
    |\Psi_i\rangle  = \alpha |ggg\rangle |00\rangle +
    \beta |egg\rangle |00\rangle = (\alpha |g\rangle +
    \beta |e\rangle ) \otimes |gg\rangle |00\rangle ,
\end{equation}
where $|\alpha|^2+|\beta|^2=1$ and the notation $|q_1\, q_2\,
q_3\rangle\, |b\, m_1\rangle$ displays the quantum states of the qubits
$Q_1$, $Q_2$, and $Q_3$, as well as the states of the bus $B$ and
memory $M_1$ resonators; the notation including the outer product sign ``$\otimes$''
uses the same order for the system elements.

{\it Step 1} of the procedure (Fig.\ 1e) is equivalent to the first
partial measurement of the qubit $Q_1$ in
Fig.\ 1a with strength $p$. This step consists of the QRQ swap
$Q_1$--$B$--$Q_2$, followed by  measurement of qubit $Q_2$. First,
the partial swap between the qubit $Q_1$ and bus $B$ with the swap
probability $p$ results in the state
\begin{equation}
    |\Psi_{1a}\rangle = \alpha |ggg\rangle |00\rangle +
    \beta e^{i\theta_p} (\sqrt{1-p} \, |egg\rangle |00\rangle
    - i e^{i\tilde\theta_{p}} \sqrt{p} \, |ggg\rangle |10\rangle) ,
\label{psi-1a}\end{equation} where $\theta_p$ and $\tilde\theta_p$
are the dynamic phases accumulated when the frequency of qubit
$Q_1$ is tuned into and out of resonance with the resonator $B$
[each term in Eq.~(\ref{psi-1a}) assumes a separate rotating
frame]. The factor $-i$ in the last term comes from the ideal
qubit-resonator evolution described by the standard Hamiltonian. After this partial swap, the resonator $B$ is no
longer in the ground state. The second part of the QRQ swap fully transfers the excitation from
$B$ into $Q_2$, resulting in the state
\begin{equation}
    |\Psi_{1b}\rangle = \alpha |ggg\rangle |00\rangle
    + \beta e^{i\theta_p} (\sqrt{1-p} \, |egg\rangle |00\rangle
    - \sqrt{p} \, e^{i\theta_{pa}} |geg\rangle |00\rangle),
  \label{psi-1b}\end{equation}
where the phase $\theta_{pa}$ combines $\tilde\theta_{p}$ and the
dynamic phase accumulated during the full swap. The minus sign in
the last term is due to the additional factor $-i$, appearing when
the excitation in the resonator $B$ swaps to $Q_2$ (this is why
the full swap is termed an ``iSWAP'').

After the QRQ swap $Q_1$--$B$--$Q_2$, the qubit $Q_2$ is measured
projectively (``strongly'') and only the outcome $|g\rangle$ is
selected. Phase qubits are measured
\cite{Hofheinz2009} by lowering the tunnel barrier between the right
and left potential wells shown in the bottom panel of Fig.\ 1a,
with a high likelihood of tunneling to the
right well if the qubit is in the excited state $|e\rangle$, while
there is a very small tunneling probability
if the qubit is in its ground state
$|g\rangle$. When a qubit that is initially in a superposition of $|g\rangle$ 
and $|e\rangle$ tunnels to the right well, the subsequent rapid energy decay 
in the right well destroys any coherence between $|g\rangle$ and $|e\rangle$ 
states. The barrier is lowered only for a few nanoseconds, and
the quantum state projection occurs during this time. Actual readout
of the measurement result
takes place many microseconds later, using a SQUID flux measurement.

In the case of the measurement result $|g\rangle$ (no tunneling for
qubit $Q_2$), the system state (\ref{psi-1b}) collapses to the state
\begin{equation}
    |\Psi_{1c}\rangle = \alpha |ggg\rangle |00\rangle
    + \beta e^{i\theta_p} \sqrt{1-p} \, |egg\rangle |00\rangle .
  \label{psi-1c}\end{equation}
Notice that while the state (\ref{psi-1b}) is normalized, $\langle
\Psi_{1b}|\Psi_{1b}\rangle =1$, the post-selected state
(\ref{psi-1c}) is not normalized, so that $\langle
\Psi_{1c}|\Psi_{1c}\rangle$ is the probability of the $|g\rangle$
outcome, while the normalized state would be
 $|\Psi_{1c}\rangle/\sqrt{\langle
\Psi_{1c}|\Psi_{1c}\rangle}$. We prefer here
to use unnormalized states as in Eq.~(\ref{psi-1c}) because these
are linearly related to the initial state, in contrast to the
normalized states. The state (\ref{psi-1c}) can be written as
$|\Psi_{1c}\rangle = (\alpha |g\rangle + \beta e^{i\theta_p}
\sqrt{1-p} \, |e\rangle ) \, |gg\rangle |00\rangle$, so at the end
of this step we essentially have a one-qubit state
in $Q_1$, even though other elements of the
system are entangled with $Q_1$ during the
evolution of this step.

{\it Step 2} of the protocol (Fig.\ 1e) involves storing $Q_1$'s
state in the memory resonator $M_1$ for a relatively long time
$\tau_2$, which corresponds to the delay $\tau$ in the protocol in Fig.~1a in the main text. We first perform an iSWAP between $Q_1$ and $M_1$, resulting in the state
\begin{equation}
    |\Psi_{2a}\rangle = \alpha |ggg\rangle |00\rangle
    -i \beta e^{i\theta_p} e^{i\tilde\theta_s} \sqrt{1-p} \,
    |ggg\rangle |01\rangle ,
   \label{psi-2a}
\end{equation}
where $\tilde\theta_s$ is the dynamic phase accumulated when tuning $Q_1$ into the resonance with $M_1$. With $Q_1$ now in its ground state, we detune $Q_1$ from $M_1$ to its ``idle'' frequency, and wait a time $\tau_2$. During this time the
state in the resonator $M_1$ decays in energy at the rate
$\Gamma=1/T_1$, where $T_1=2.5 \, \mu$s is the energy relaxation
time of $M_1$, so that the overall
decay factor is $\kappa_2=e^{-\Gamma \tau_2}$ (pure dephasing is negligible). 

The decay in $M_1$ can be treated by considering two scenarios: \cite{Korotkov2010} either the state of $M_1$ ``jumps''
to $|g\rangle$ during the storage time $\tau_2$ or there is no jump.
In the jump scenario the resulting unnormalized state is
    \begin{equation}
    |\Psi_{2b}^j\rangle = \beta \sqrt{1-p}\,
    \sqrt{1-e^{-\Gamma\tau_2}} \, |ggg\rangle |00\rangle,
    \label{psi-2b-j}\end{equation}
where the overall phase is not important. We will return to this
scenario later, focusing first on the no-jump scenario, which
produces the unnormalized state
\begin{equation}
    |\Psi_{2b}^{nj}\rangle = \alpha |ggg\rangle |00\rangle
    -i \beta e^{i\theta_p} e^{i\tilde\theta_s} \sqrt{1-p} \,
    e^{-\Gamma\tau_2/2}
    |ggg\rangle |01\rangle .
\end{equation}

After the storage time $\tau_2$ we swap the state in $M_1$ back to $Q_1$, so that
at the end of step 2 the no-jump state becomes
\begin{equation}
    |\Psi_{2c}^{nj}\rangle = \alpha |ggg\rangle |00\rangle
    + \beta e^{i(\theta_p+\theta_s)} \sqrt{1-p} \, e^{-\Gamma\tau_2/2}
    |egg\rangle |00\rangle ,
   \label{psi-2c-nj}
\end{equation}
where the phase $\theta_s$ includes $\tilde\theta_s$ [see Eq.\
(\ref{psi-2a})], the similar dynamic phase accumulated during the
swap back to $Q_1$, the $\pi$-shift due to the factor $(-i)^2$, and
the phase $2 \pi \Delta f \tau_2$ accumulated due to the frequency
difference $\Delta f$ between the resonator $M_1$ and the qubit
$Q_1$ at its ``idle'' frequency. After the step is completed, we again
have essentially a one-qubit state.

{\it Step 3} of the protocol consists of a $\pi_x$ rotation,
the second QRQ swap $Q_1$--$B$--$Q_3$ with strength $p_u$, and the
projective measurement of $Q_3$ (this step is analogous to the
second partial measurement in Fig.\ 1a). The $\pi_x$ rotation
applied to $Q_1$ exchanges the amplitudes of its $|g\rangle$ and
$|e\rangle$ states in Eq.~(\ref{psi-2c-nj}):
\begin{equation}\label{eq1}
    |\Psi_{3a}^{nj}\rangle = \alpha |egg\rangle |00\rangle +
    \beta e^{i(\theta_p +\theta_s)} \sqrt{1-p} \,
    e^{-\Gamma\tau_2/2}  |ggg\rangle |00\rangle .
\end{equation}
The partial swap between $Q_1$ and $B$ then yields the state
\begin{eqnarray}
    && |\Psi_{3b}^{nj}\rangle = \alpha e^{i\theta_u}(\sqrt{1-p_u} \,
    |egg\rangle |00\rangle - i e^{i\tilde\theta_u} \sqrt{p_u} \,
    |ggg\rangle |10\rangle) \hspace{1.2cm} \nonumber \\
    && \hspace{1.2cm} + \beta e^{i(\theta_p+\theta_s)}
        \sqrt{1-p} \,  e^{-\Gamma\tau_2/2}  |ggg\rangle |00\rangle,
    \label{eq2}
\end{eqnarray}
where $\theta_u$ and $\tilde\theta_u$ are the dynamic phases
accumulated during this partial swap. Next, the QRQ swap is
completed with a full iSWAP between $B$ and $Q_3$, yielding the
state
\begin{eqnarray}
    &&    |\Psi_{3c}^{nj}\rangle = [(e^{i\theta_u} \alpha \sqrt{1-p_u}
    \,  |egg\rangle + e^{i(\theta_p+\theta_s)} \beta \sqrt{1-p} \,
    e^{-\Gamma\tau_2/2} |ggg\rangle )
    \nonumber \\
    && \hspace{1.2cm}                - e^{i(\theta_u + \theta_{ua})}
    \alpha \sqrt{p_u} \, |gge\rangle]  \otimes |00\rangle ,
    \label{psi-3c-nj}
\end{eqnarray}
where $\theta_{ua}$ combines $\tilde\theta_u$ and the dynamic phase
accumulated during the last iSWAP. Finally, the measurement of $Q_3$
and the selection of the result $|g\rangle$ (thus corresponding to an overall double-null
outcome) produces the no-jump state
\begin{equation}
  |\Psi_{f}^{nj}\rangle = (\alpha \sqrt{1-p_u}
    \,  |e\rangle + e^{i(\theta_p+\theta_s-\theta_u)} \beta \sqrt{1-p}
    \, e^{-\Gamma\tau_2/2} |g\rangle ) \otimes |gg\rangle |00\rangle ,
    \label{psi-f-nj}
\end{equation}
where we ignore the unimportant overall
phase.

Equation (\ref{psi-f-nj}) coincides with Eq.~(3) of the main
text, if we neglect the dynamic phase $\theta_p+\theta_s-\theta_u$.
This phase does not depend on the initial state, but in general
depends on $p$, $p_u$, and $\tau_2$. To restore the initial qubit
state (up to a $\pi_x$ rotation), this phase can be corrected by an
additional single-qubit phase gate (rotation about the $z$ axis of the
Bloch sphere). In the experiment we typically did not perform this
correction, and instead compensated for this phase numerically in the
quantum process tomography analysis. However, we have checked
explicitly that for the initial states $\ket{g}-i\ket{e}$ and
$\ket{g}+\ket{e}$ (using the same QED protocol parameters), the
measured output states differ by a phase of $\pi/2$, as expected.

Note that we
completely omit step 2 when testing the protocol with no storage in
$M_1$, i.e. with $\tau_2=0$ (see Fig.\ 2 of the main text).
  In this
case there is no dynamic phase $\theta_s$ in Eq.~(\ref{psi-f-nj}),
we have no delay-based decay so that $e^{-\Gamma\tau_2/2} = 1$, and
also the dynamic phases $\theta_p$ and
$\theta_u$ cancel each other because $p_u=p$ and therefore
$\theta_u=\theta_p$. In reality there is
still a small amount of energy decay occurring in steps 1 and 3. We
take this into account in the numerical simulations as described in
the next section.

Now let us return to the scenario when the energy relaxation
event (the jump) occurs  during step 2, producing the state
$|\Psi_{2b}^j\rangle$ given by Eq.~(\ref{psi-2b-j}). After performing the swap between the memory resonator and $Q_1$, this state remains the same, $  |\Psi_{2c}^j\rangle=  |\Psi_{2b}^j\rangle$, because all elements are in their ground states. In step 3 of the
protocol, following the $\pi_x$ pulse, the state becomes
\begin{equation}
    |\Psi_{3a}^j\rangle = \beta \sqrt{1-p}\,
    \sqrt{1-e^{-\Gamma\tau_2}} \, |egg\rangle |00\rangle ,
    \label{psi-3a-j}
\end{equation}
and following the partial swap between $Q_1$ and $B$ this state evolves
into
\begin{equation}
    \hspace{-0.5cm}  |\Psi_{3b}^{j}\rangle = \beta \sqrt{1-p}\,
    \sqrt{1-e^{-\Gamma\tau_2}} \,
    (\sqrt{1-p_u} \,
    |egg\rangle |00\rangle - i e^{i\tilde\theta_u} \sqrt{p_u} \,
    |ggg\rangle |10\rangle)
    \label{psi-3b-j}
\end{equation}
(the overall phase $\theta_u$ is now unimportant), and after the
full iSWAP between $B$ and $Q_3$ it becomes
\begin{equation}
    \hspace{-0.5cm}  |\Psi_{3c}^{j}\rangle = \beta \sqrt{1-p}\,
    \sqrt{1-e^{-\Gamma\tau_2}} \,
    (\sqrt{1-p_u} \,  |egg\rangle
     -  e^{i\theta_{ua}}
    \sqrt{p_u} \,
    |gge\rangle )  \otimes |00\rangle .
    \label{psi-3c-j}
\end{equation}
After the measurement of $Q_3$ and selection of the null result
$|g\rangle$, the final state in the jump scenario is
\begin{equation}
    \hspace{-0.5cm}  |\Psi_{f}^{j}\rangle = \beta \sqrt{1-p}\,
    \sqrt{1-e^{-\Gamma\tau_2}} \,
    \sqrt{1-p_u} \,  |e\rangle \otimes |gg\rangle |00\rangle ,
    \label{psi-f-j}
\end{equation}
so that the qubit $Q_1$ is now in the $|e\rangle$ state.

The squared norm of the no-jump final state
$|\Psi_{f}^{nj}\rangle$ in Eq.~(\ref{psi-f-nj}) is the
probability of the no-jump scenario (which includes the double-null
outcome selection),
\begin{equation}
    P^{nj}_f \equiv \langle \Psi_{f}^{nj}|\Psi_{f}^{nj}\rangle =
    |\alpha|^2 (1-p_u) +|\beta|^2 (1-p)e^{-\Gamma\tau_2}.
\end{equation}
Notice that this probability becomes $P^{nj}_f=(1-p)e^{-\Gamma\tau_2}$ if we
choose $1-p_u=(1-p)e^{-\Gamma\tau_2}$. The squared norm of the state
$|\Psi_{f}^{j}\rangle$ in Eq.~(\ref{psi-f-j}) is the probability of
the jump scenario,
\begin{equation}
    P^{j}_f \equiv \langle \Psi_{f}^{j}|\Psi_{f}^{j}\rangle =
    |\beta|^2 (1-p) (1-p_u) (1-e^{-\Gamma\tau_2}).
\end{equation}
This probability is given by
$P_f^j=|\beta|^2 (1-p)^2 e^{-\Gamma\tau_2}(1-e^{-\Gamma \tau_2}) $
if we choose $1-p_u=(1-p)e^{-\Gamma\tau_2}$.
The probabilities $P_{f}^{nj}$ and $P_f^j$
cover all possible double-null outcomes in
this model, so their sum
\begin{equation}
    P_{DN} = P_{f}^{nj}+P_{f}^j
\end{equation}
is the probability of the 
double-null outcome.

Combining the two scenarios, the normalized density matrix
of the system after the selection of the double-null outcome is
\begin{equation}
    \rho_{f} =\frac{ |\Psi_{f}^{nj}\rangle \langle \Psi_{f}^{nj}|
    + |\Psi_{f}^{j}\rangle \langle \Psi_{f}^{j}|}{P_{DN}}.
\end{equation}

In this double-null outcome, note that the target
qubit $Q_1$ is now unentangled with the other elements, which are all in their ground states. Comparing the resulting state of the qubit
$Q_1$ with the corresponding final state in the single-qubit
protocol based on partial tunneling [see Fig.\ 1a and Eq.~(1) in
the main text], we see only two differences: the non-zero dynamic phase
$\theta_p+\theta_s-\theta_u$ in Eq.~(\ref{psi-f-nj}), and the exchange
of the amplitudes of the states $|g\rangle$ and $|e\rangle$ due to the absence of the
final $\pi_x$ pulse. Therefore, our experimental protocol shown in
Fig.\ 1e essentially realizes the un-collapsing protocol shown in
Fig.\ 1a, but with much better experimental fidelity.

\section{Numerical simulations}

For numerical simulations we follow the theory of Ref.\
\cite{Korotkov2010} and describe decoherence by the energy
relaxation factors $\kappa_1$, $\kappa_2$, and $\kappa_3$
(each factor for the corresponding step of
the protocol shown in Fig.\ 1e) and by the factor $\kappa_\varphi$,
which accounts for pure dephasing during the whole procedure.
 The primary decay factor
is $\kappa_2 \approx \exp(-\tau_2/T_1)$, where $\tau_2=\tau$ is the
storage time and $T_1=2.5 \, \mu$s is the energy relaxation time of
the memory resonator. Similarly, $\kappa_1$ describes energy
relaxation before the first partial measurement and $\kappa_3$
describes energy relaxation in between the $\pi_x$ pulse and the
second partial measurement. Therefore $\kappa_1=\exp(-\tilde
\tau_1/T_1^{(1)})$ and $\kappa_3=\exp(-\tilde\tau_3/T_1^{(3)})$,
where $\tilde\tau_1$ is the effective duration of  step 1 in Fig.\
1e before the quantum information is partially swaped into the bus
resonator, $\tilde\tau_3$ is the effective duration of step 3
between the $\pi_x$ pulse and partial swap into the bus resonator,
and $T_1^{(1)}$ and $T_1^{(3)}$ are the effective energy relaxation
times for these steps (mostly determined by the phase qubit $Q_1$).
We estimate that $\kappa_1 \approx \kappa_3 \approx 0.985$, consistent
with the energy relaxation time $T_1\simeq 0.6\ \mu$s of the phase
qubit $Q_1$ (see Table \ref{tab1}) and the time $\sim 10$ ns, which
the quantum state spends in the phase qubit before the first partial
swap (in step 1) and between the $\pi_x$ pulse and the second
partial swap (in step 3).

The overall pure dephasing factor is
$\kappa_\varphi=\exp[-\tau_1/T_\varphi^{(1)}-\tau_2/T_\varphi^{(2)}
-\tau_3/T_\varphi^{(3)}]$, where $T_\varphi^{(i)}$ is the effective
pure dephasing time during $i$th step ($1/T_\varphi=1/T_2-1/2T_1$).
In simulations we used the value $\kappa_\varphi=0.95$, which fits
well with the experimental results and is consistent with the qubit
parameters in Table \ref{tab1}. Notice that $T_\varphi^{(2)}$ is
very long since during step 2, the quantum state is stored in the
memory resonator, and therefore $\kappa_{\varphi}$ does not depend
on $\tau_2$. Also notice that because of the $\pi_x$ pulse in the
procedure (see Fig.\ 1e), pure dephasing is reduced,
essentially due to a spin-echo effect. In
the theory we neglect imperfections of the unitary gates and the
qubit decoherence after the second partial swap; we also do not
accurately consider decoherence processes in the actual
multi-component device, essentially reducing it to the single-qubit
model of Ref.\ \cite{Korotkov2010}. In a
practical sense, however, these additional imperfections are
somewhat accounted for by small
 adjustments of the parameters $\kappa_1$,
$\kappa_3$, and $\kappa_\varphi$. We have checked numerically that
slight variations of the parameters $\kappa_1$, $\kappa_3$, and
$\kappa_\varphi$ do not affect the simulation results significantly;
$\kappa_3$ is the most important parameter,
and varying its value in the
experimentally-expected range of $0.985\pm 0.005$ gives good
agreement with the data shown in Fig.\ 3a of the main text.

In the experiment we do not perform the final $\pi_x$ rotation to
save time, so in the final state the amplitudes of the states
$\ket{g}$ and $\ket{e}$ are exchanged in comparison
with the initial state
$|\psi_{i}\rangle=\alpha\ket{g}+\beta\ket{e}$ in $Q_1$ (here and below we use a lowercase $|\psi\rangle$ to represent the state of $Q_1$, in contrast to $|\Psi\rangle$ which represents the state of the complete system of 3 qubits and 2 resonators). Following the approach of Ref.\ \cite{Korotkov2010},
neglecting the dynamic phases, and for the moment neglecting pure
dephasing, we can represent the state of the qubit $Q_1$
after the double-null outcome selection as an incoherent mixture
of the three states $|g\rangle$, $|e\rangle$, and
\begin{equation}
    |\psi_f^{nj}\rangle =\beta\sqrt{\kappa_1\kappa_2(1-p)}\ket{g}
    +\alpha\sqrt{\kappa_3(1-p_u)}\ket{e}.
    \label{psi-nj}
\end{equation}
The unnormalized state $|\psi_f^{nj}\rangle$ occurs
in the ``no jump'' scenario
during steps 1, 2, and 3. The squared norm of this wavefunction is
the probability of the no-jump scenario,
\begin{equation}
    P_f^{nj}= \langle \psi_f^{nj} | \psi_f^{nj} \rangle =
    \left|\alpha\right|^2\kappa_3(1-p_u)
    +\left|\beta\right|^2\kappa_1\kappa_2(1-p),
\end{equation}
which includes the probability of the double-null outcome selection.

The final state $|g\rangle$ is realized if there was a ``jump'' to
$|g\rangle$ after the  $\pi_x$ pulse in step 3 and there was zero or one jump during steps 1 and 2.  This occurs with the probability
\begin{equation}
    P_f^{\ket{g}}=(1-\kappa_3)\left|\alpha\right|^2
    +(1-\kappa_3)\left|\beta\right|^2[(1-\kappa_1)
    +\kappa_1(1-p)(1-\kappa_2)],
\end{equation}
which can be easily understood in the classical way (for a qubit
starting either in the state $|g\rangle$ or $|e\rangle$). The final state
$|e\rangle$ is realized if there was a jump either during step 1 or
2 and no jump during step 3; this occurs with probability
\begin{equation}
    P_f^{\ket{e}}=\left|\beta\right|^2[(1-\kappa_1)
    +\kappa_1(1-p)(1-\kappa_2)]\kappa_3(1-p_u).
\end{equation}

Combining these three scenarios, we obtain the normalized
density matrix of the qubit final state:
\begin{equation}
  \rho_f = \frac{|\psi_{f}^{nj}\rangle\langle \psi_{f}^{nj}|
  + P_f^{\ket{g}} |g\rangle\langle g |+
  P_f^{\ket{e}} |e\rangle\langle e|  }{P_{DN}} , \,\,\,\,\,
    \label{rho-f-sum}
\end{equation}
where
\begin{equation}
     P_{DN}=P_{f}^{nj}+P_f^{\ket{g}}+P_f^{\ket{e}}
\end{equation}
is the probability of the double-null outcome. Notice that there is
no factor $P_{f}^{nj}$ in the numerator of Eq.~(\ref{rho-f-sum})
because it was included in the definition of the unnormalized
state $|\psi_{f}^{nj}\rangle$ in Eq.~(\ref{psi-nj}). The
unnormalized final density matrix $P_{DN}\rho_f$ [the numerator in
Eq.~(\ref{rho-f-sum})] is linearly related  to the initial density
matrix $\rho_i=|\psi_i\rangle\langle\psi_i |$, so the linear map
used in the analysis of the quantum process tomography is
$\rho_i\rightarrow P_{DN}\rho_f$.

Pure dephasing (described by $\kappa_\varphi$) does not affect the
probabilities and does not affect the final states $|g\rangle$ and
$|e\rangle$. The only effect of pure
dephasing is that the off-diagonal matrix elements of
$|\psi_{f}^{nj}\rangle\langle \psi_{f}^{nj}|$ are multiplied by
$\kappa_\varphi$. This is equivalent to multiplying the off-diagonal
matrix elements of $\rho_f$ given by Eq.~(\ref{rho-f-sum}) by
$\kappa_\varphi$. In other words, pure dephasing can be thought of
as occurring after (or before) the procedure described by Eq.~(\ref{rho-f-sum}).

The dynamic phases appearing in the actual experimental
procedure affect only the relative phase between the two terms in
Eq.~(\ref{psi-nj}). Therefore, the dynamic phases can be taken into
account by using a single parameter: the phase shift of the
off-diagonal element of the final density matrix. This dynamic phase
shift depends on the parameters of the experimental protocol,
including the strength $p$ and $p_u$ of the two partial measurements
(partial swaps) and the storage duration $\tau_2$.

\section{Analysis of the QED process fidelity}

In the main text we use the definition \cite{Kiesel2005}
\begin{equation}
    {\mathcal F}= \frac{\Tr(\chi^{\textrm{ideal}} \chi)}{\Tr(\chi)},
    \label{fid-def}
\end{equation}
for the process fidelity of a non-trace-preserving quantum
operation. This definition implies that  $\chi/\Tr(\chi)$ is the
effective process matrix (which is shown e.g. in Fig.\ 2a of the
main text). Notice that $\chi/\Tr(\chi)$ does not correspond to any
physical trace-preserving process; however, this is a positive
Hermitian matrix with unit trace, and therefore $0\leq {\mathcal
F}\leq 1$ when $\chi^{\textrm{ideal}}$ corresponds to a unitary
operation. The perfect fidelity, ${\mathcal F}=1$, requires $\chi=
P_{\rm s}  \chi^{\textrm{ideal}}$ with $P_{\rm s} \leq 1$ being the
selection probability (in this case $P_{\rm s}$ should not depend on
the initial state). This justifies the definition (\ref{fid-def}).

However, Eq.~(\ref{fid-def}) is not the only possible definition
for the fidelity of a non-trace-preserving quantum process. For
example, another natural definition~\cite{Korotkov2010} is the
averaged state fidelity,
\begin{equation}\label{eqS1}
    {\mathcal F}_{\rm av} = \frac{\int \Tr(\rho_{f}\rho^{\rm{ideal}}_{f})
    \,  d|\psi_i\rangle }{\int d|\psi_i\rangle} ,
\end{equation}
where $\rho_{f}^{\rm ideal}=U|\psi_{i}\rangle \langle
\psi_{i}|U^\dagger$, $U$ is the desired unitary operation, $\rho_f$
is the actual normalized density matrix, and the integration is over
all pure initial states $|\psi_i\rangle$ with uniform weight (using
the Haar measure); in the one-qubit case this is the uniform
averaging over the Bloch sphere. Another natural definition is the
averaged state fidelity, which is averaged with a weight
proportional to the selection probability $P_{\rm s}$ (denoted
$P_{\rm DN}$ in the main text),
\begin{equation}\label{eqS2}
    {\mathcal F}'_{\rm av} = \frac{\int\Tr(\rho_{f}
\rho^{\rm{ideal}}_{f}) P_{\rm s}(|\psi_i\rangle ) \, d|\psi_i\rangle
} {\int P_{\rm s}(|\psi_i\rangle ) \, d|\psi_i\rangle} .
\end{equation}
Notice that both ${\mathcal F}_{\rm av}$ and ${\mathcal F}'_{\rm
av}$ can be easily calculated when the process matrix $\chi$ is
known.

\begin{figure*}
  \centering
  \includegraphics[width=5.7in,clip=True]{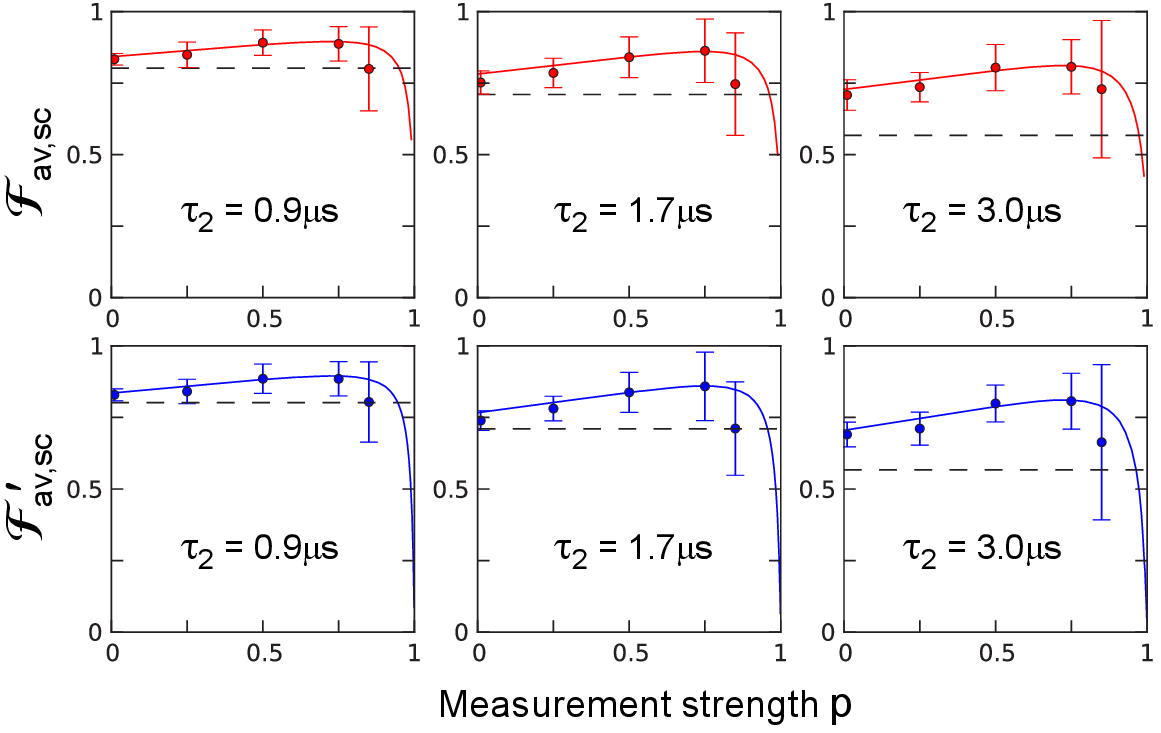}
  \caption{\label{figS1}\footnotesize{\textbf{QED process fidelity
  characterized using different methods.}
For the same experimental data as in Fig.\ 3a of the main text,
here we show the QED process fidelities calculated via the averaged
state fidelities ${\mathcal F}_{\rm av}$ and ${\mathcal F}_{\rm
av}'$ defined in Eqs.\ (\ref{eqS1}) and (\ref{eqS2}). For easier
comparison with Fig.\ 3a we show the scaled results ${\mathcal
F}_{\rm av, sc}=(3{\mathcal F}_{\rm av}-1)/2$ (top panels) and
${\mathcal F}_{\rm av, sc}'=(3{\mathcal F}_{\rm av}'-1)/2$ (bottom
panels). As in Fig.\ 3a, the quantum state is stored for the
durations $\tau_2 = 0.9,~1.7$, and 3 $\mu$s in the memory resonator
$M_1$, which has the energy relaxation time $T_1=2.5\, \mu$s. The
measurement strength (swap probability) $p$ is indicated on the
horizontal axis, and the uncollapsing swap probability $p_u$ is
adjusted as described in the main text.
  Circles with error bars are the experimental results; lines are
simulations. Horizontal dashed lines in each panel show the
free-decay process fidelity. The statistical errors increase with
increasing measurement strength $p$ due to the decrease in sample
size (fewer double-null outcomes). It is seen that all definitions
of the QED process fidelity give similar results, and all of them
show significant increase of the storage fidelity compared with the
case of natural energy relaxation.
  }
  }
\end{figure*}

For a trace-preserving quantum operation ${\mathcal F}'_{\rm av}=
{\mathcal F}_{\rm av}$ because $P_{\rm s}=1$, and there is a direct
relation \cite{Nielsen2002} $\Tr(\chi^{\textrm{ideal}}\chi ) = [(d
+ 1){\mathcal F}_{\rm av}-1]/d$, where $d$ is the dimension of the
Hilbert space ($d=2$ in our one-qubit case). It is possible to show
that in the general non-trace-preserving case the same relation
remains valid between ${\mathcal F}$ defined by Eq.~(\ref{fid-def})
and ${\mathcal F}_{\rm av}'$ defined by Eq.~(\ref{eqS2}),
\begin{equation}
    {\mathcal F} =\frac{(d + 1)\,{\mathcal F}_{\rm av}'-1}{d}.
    \label{rel-F-Fav'}
\end{equation}
Notice that the denominator ${\Tr (\chi)}$ in Eq.~(\ref{fid-def})
is equal to the averaged selection probability,
\begin{equation}
    {\Tr (\chi)}=\frac{\int P_{\rm s}(|\psi_i\rangle ) \, d|\psi_i\rangle} {\int
    d|\psi_i\rangle}.
\end{equation}

We have numerically calculated the process fidelity in our
un-collapsing QED experiment using all three definitions
(\ref{fid-def})--(\ref{eqS2}). For easier comparison with the
results for ${\mathcal F}$ shown in Fig.\ 3a of the main text, in
Fig.~\ref{figS1} we scale ${\mathcal F}_{\rm av}$ and ${\mathcal
F}_{\rm av}'$ as in Eq.~(\ref{rel-F-Fav'}): ${\mathcal F}_{\rm av,
sc}=(3{\mathcal F}_{\rm av}-1)/2$, ${\mathcal F}_{\rm av,
sc}'=(3{\mathcal F}_{\rm av}'-1)/2$. Notice that for the
experimental results ${\mathcal F}_{\rm av, sc}'$ and ${\mathcal F}$
are not exactly equal to each other [in spite of Eq.\
(\ref{rel-F-Fav'})] because slightly different algorithms were used
in the numerical processing of the over-complete experimental data
set. Comparing Fig.~\ref{figS1} with Fig.\ 3a, we see that the
experimental results using the three fidelity definitions are close
to each other, and all of them show significant increase of the
fidelity due to the un-collapsing-based QED procedure.

\end{document}